\documentclass[english, letterpaper, 12pt, fleqn]{article}

\usepackage[T1]{fontenc}
\usepackage[utf8]{inputenc}
\usepackage{babel}
\usepackage{amsmath}
\usepackage{amssymb}
\usepackage{bm}
\usepackage{mathtools}
\usepackage{graphics}
\usepackage{graphicx}
\usepackage{epsfig}
\usepackage{fullpage}

\makeatletter
\@addtoreset{chapter}{part}
\makeatother

\newcommand\be{ \begin{equation} }
\newcommand\ee{\end{equation}}
\newcommand\bea{\begin{eqnarray} \nonumber }
\newcommand\eea{\end{eqnarray}}

\DeclareMathOperator{\e}{e}

\newcommand{\bi}{\begin{itemize}}
\newcommand{\ei}{\end{itemize}}

\newcommand{\given}{\:\vert\:}
\newcommand{\bgiven}{\:\Big\vert\:}

\begin{document}

\title{Tail protection for long investors: Trend convexity at work}

\author{T.-L. Dao, T.-T. Nguyen, C. Deremble, Y. Lemp\'eri\`ere,  J.-P. Bouchaud, M. Potters}
\date{\today}
\maketitle

\abstract{
The performance of trend following strategies can be ascribed to the difference between long-term and short-term realized variance. 
We revisit this general result and show that it holds for various definitions of trend strategies. This explains the positive convexity 
of the aggregate performance of Commodity Trading Advisors (CTAs) which -- when adequately measured -- turns out to be much stronger than anticipated. 
We also highlight interesting connections with so-called Risk Parity portfolios. Finally, we propose a new portfolio of strangle options that provides a pure exposure to the long-term variance of the underlying, 
offering yet another viewpoint on the link between trend and volatility.
}

\newpage
\tableofcontents
\newpage

\section{Introduction}

A key concept in finance is the idea of dynamic asset allocation: by adequately rebalancing a portfolio, one can effectively transform a simple linear exposure to an underlying to a much more complex pay-off profile, 
in some cases effectively hedging away part of the risks \cite{FB_RJ,PE_SH}. Since a majority of investors have a long exposure to stock markets, a strategy that allows one to mitigate the losses when the market goes down 
sounds like a very useful idea, and, if available, should be highly valued by those investors. As has been already pointed out several times, trend following strategies appear to offer such a downside protection. The aim of the present paper is to revisit this important theme, and hopefully add both some interesting new insights and convincing empirical illustrations.

\begin{figure}
\begin{center}
\epsfig{file=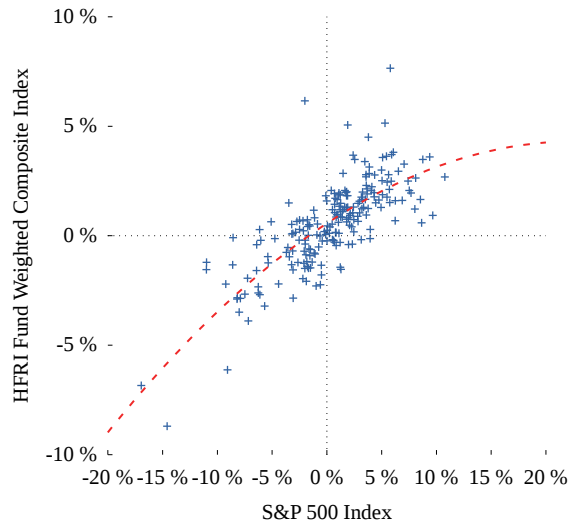}
\caption{Monthly returns of the HFRI Fund Weighted Composite Index vs. monthly returns of the S\&P 500 Index. We see a concave relationship: hedge funds appear to do worse when volatility is high. }
\label{HFRIconv}
\end{center}
\end{figure}

The obvious hedge against market drops is simply being long index options. Unfortunately, options are usually sold at a premium, meaning that the average performance of these long
volatility portfolios is strongly negative (with occasional rallies during crises). This protection works by design, but defeats the purpose since most of the expected gains from the market are erased by its cost. 

Hedge funds seem to offer yet another alternative. The hedge fund industry often claims that its performance is not correlated to that of the market, and therefore provide a useful source of diversification, as well as 
valuable ``alpha'' (positive average performance on top of the market). However, if one plots the monthly performance of a global hedge fund index (the HFRI Fund Weighted Composite Index\footnote{Data from Hedge Fund Research (HFR)}) as a function of the contemporaneous market return, as we do in figure \ref{HFRIconv}, one sees not only a strong positive correlation between the two (i.e. when the market goes down, so do hedge funds returns), but also a {\it negative convexity}. In other words, the performance appears to
become worse than average in periods of high market volatility: at least in aggregate, hedge funds have difficulty fulfilling their promises in terms of tail diversification 
(see \cite{FW_HD} for a detailed analysis of hedge fund performance). A related observation was made in \cite{YL_RP}, where it was shown that the performance of the HFRI Index has a significant skewness, making its performance akin to that of shorting volatility.

\begin{figure}
\begin{center}
\epsfig{file=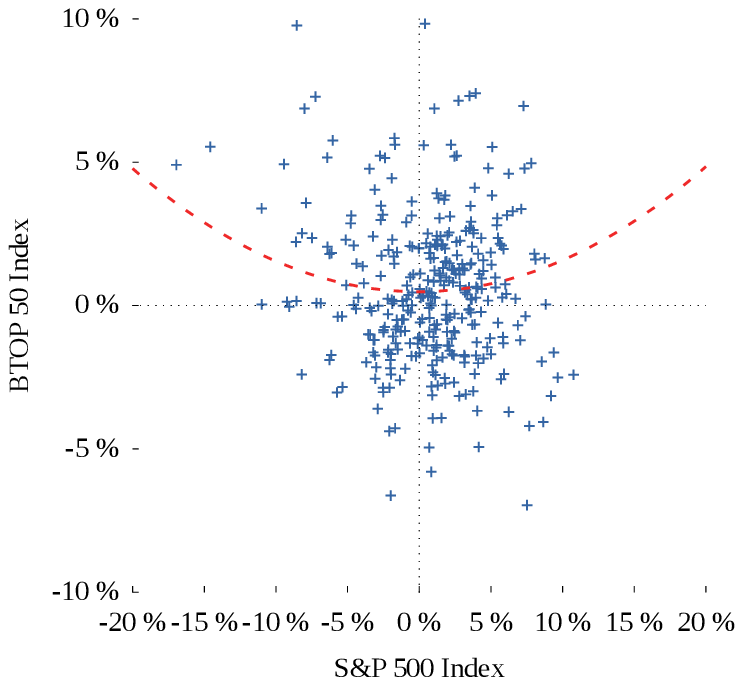}
\epsfig{file=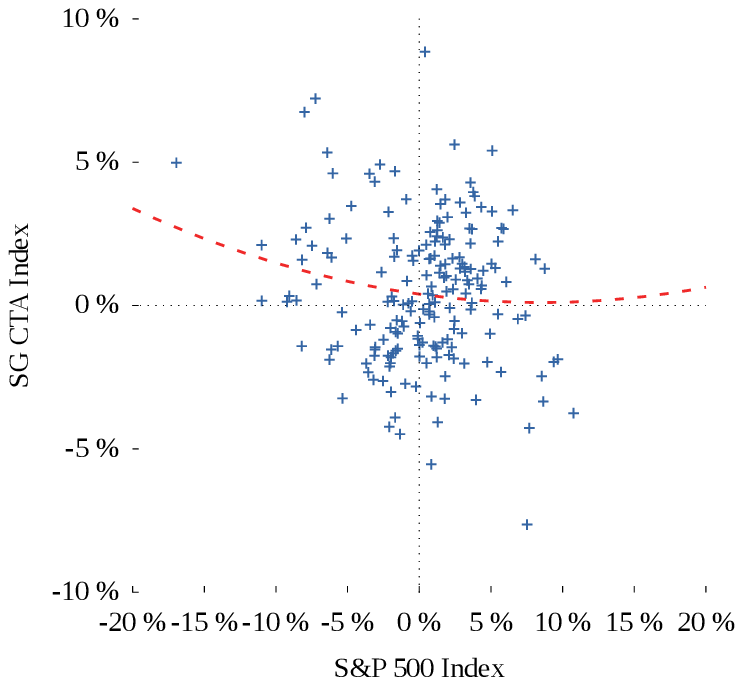}
\caption{Left: Monthly returns of the Barclays BTOP 50 Index vs. monthly returns of the S\&P 500 Index (symbols) and parabolic fit (dashed line). The convexity is hardly visible, with a low $R^2$ ($R^2=0.02$).
Right: Monthly returns of the SG CTA Index vs. monthly returns of the S\&P 500 Index (symbols) and parabolic fit (dashed line), with again $R^2=0.02$.
}
\label{BTOPconv}
\end{center}
\end{figure}

As also emphasized in \cite{YL_RP}, one interesting exception is the trend following strategy followed by CTAs. As a case in point, their performance during the 2008 Lehman crisis was very strong, 
triggering a subsequent massive rise in assets (current estimates are in the range of USD 300bn, see \cite{MUNDT}). This has been confirmed by various studies who have looked at CTA performance during or immediately after market crashes, and found above average performances \cite{FW_HD3, RM_AB} (apparently followed by below-par results in lower volatility periods, see \cite{MH_JB} and references therein). We illustrate these conclusions in Fig. \ref{BTOPconv}, where we plot the monthly returns of the two major CTA indices (Barclays' BTOP 50 and the SG CTA Index) again as a function of the contemporaneous market return. Although noisy, these plots suggest that such strategies have indeed performed better (on average) when the market volatility was high. Unfortunately, this positive convexity is hardly detectable, as quantified by the very low value of the $R^2 \approx 0.02$ for both quadratic fits shown in Fig. \ref{BTOPconv} (see however \cite{FW_HD2}). We want to be sure that we are not looking at a statistical artifact. The aim of this paper is to understand the mechanism at work behind the convex behaviour of CTAs' performance, and to find a better way to quantify it, such that we can reasonably be sure that this property is not a statistical fluke. This will be confirmed much more convincingly by our figure \ref{NEIconvexity} below.

As first noticed in \cite{FW_HD} and again demonstrated below, the performance of CTAs can indeed be mostly explained by one single strategy: trend following \cite{NB_RK}.
The algorithmic nature of trend following makes it very suitable to the quant trading style favoured by many CTA funds \cite{COV}.
Its robustness and stability to a wide range of parameters, the fact that it works across a large set of asset classes, and its undisputably positive out-of-sample performance in the last 20 years 
makes this strategy quite remarkable. In a previous paper, our group has established the universality and persistence of this effect over 200 years \cite{YL_TR} (see also \cite{AQR,WIN}),
making it one of the most significant market anomalies ever documented (barring high frequency effects, although the latter are in fact quickly eroding). 
Here, we want to revisit the convexity inherent to any trend following strategy (see, e.g. \cite{} for previous works) and provide new tools to elicit and quantify this convexity on empirical data. 

We start by deriving some general equations relating the performance of a single-asset trend following strategy to the difference between long-term and short-term realized variance. We show that this result is rather general, and holds for various definitions of the trend. We then prove that this single-asset trend shows the expected convexity properties, and provide the adequate tools to reveal this convexity on empirical data. To understand the performance of the CTA industry, we replicate the SG CTA Index\footnote{Data available at the following URL: \url{http://www.barclayhedge.com/research/indices/calyon}}, using surprisingly few parameters. Using an appropriate measure of convexity, we will elicit a much stronger effect than what is seen in the naive plot on figure \ref{BTOPconv}. We will then investigate in detail the impact of diversification on this convexity, and find interesting connections with Risk Parity strategies. Finally, we revisit the simplest convex strategy: buying options. We consider a portfolio of strangle options that provides a pure, model-free exposure to the long-term variance of the underlying. We find that the pay-off of our newly defined variance-swap is strikingly similar to that of a simple trending strategy. This construction sheds more light on the link between trend and volatility, somehow justifying the ``long vol" attribute of trend strategies. It also helps us understand the differences between long-option portfolios and trending, and the exact role of hedging.

\section{The trend on a single asset}

\subsection{Volatility on different time scales: the signature plot}

Let us start by recalling simple results about the volatility of correlated random walks. Suppose the price at time $t$, $S_t$, is written as the sum over past price changes $D_{t'<t}$:\footnote{Note that
we will consider additive random walk models for the price. All the results of this paper can however be easily extended to multiplicative random walks, i.e. considering $S_t$ to be the logarithm of the price, 
rather than the price. On the time scales we will be interested in, this distinction is however irrelevant in most practical cases.}
\be
S_t = S_{0} + \sum_{t'=1}^{t} D_{t'}  \:\text{.}
\ee
We assume that the sequence of price changes $D_{t'<t}$ are stationary random variables with zero mean and covariance given by:
\be
\label{correl}
\langle D_{u} D_{v} \rangle = \mathcal{C}(|u-v|)  \:\text{,}
\ee
where here and henceforth, $\langle ... \rangle$ denotes averaging. The case of uncorrelated random walks corresponds to $\mathcal{C}(u)=\sigma^2 \delta_{u,0}$. Trending random walks are such that $\mathcal{C}(u > 0) > 0$, while 
mean-reverting random walks are such that $\mathcal{C}(u > 0) < 0$. How does this translate in terms of the volatility of the walk?

We define the volatility of scale $\tau$ as usual:
\be
\sigma^2(\tau) := \frac{1}{\tau} \left\langle \left( S_{t+\tau} - S_t \right)^2 \right\rangle  \:\text{,}
\ee
such that $\sigma^2(1)=\sigma^2$. The exact formula for $\sigma^2(\tau)$ in terms of $\mathcal{C}(u)$ is easy to derive and reads:
\be
\sigma^2(\tau) = \sigma^2 +  \frac{2}{\tau} \sum_{u=1}^\tau \left(\tau-u \right) \, \mathcal{C}(u) \:\text{.}
\ee

\begin{figure}
\begin{center}
\epsfig{file=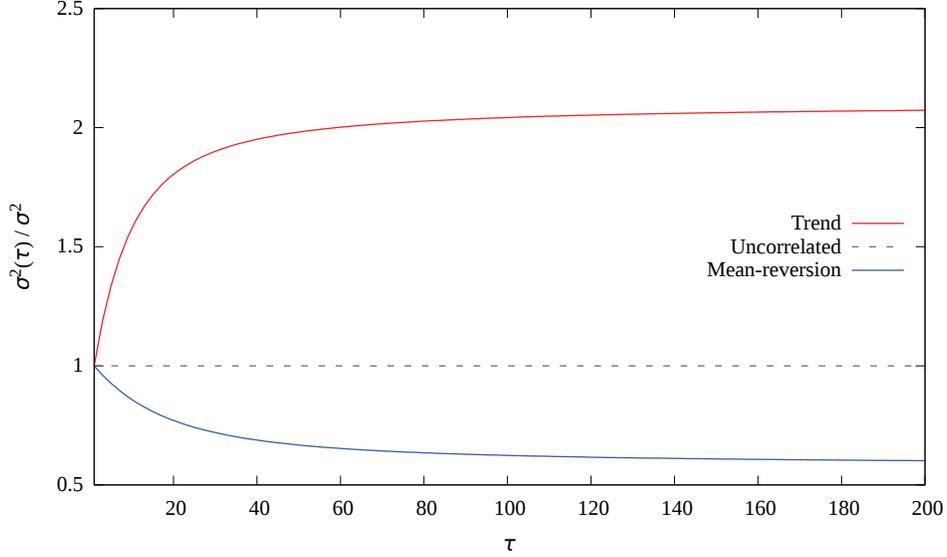}
\caption{Volatility as a function of scale $\tau$. Red line: positive autocorrelation (trends), $\mathcal{C}(u>0) = 0.1 \e^{\frac{1 - u}{5}}$.
Dashed line: uncorrelated random walk.
Blue line: negative autocorrelation (mean-reversion), $\mathcal{C}(u) = -0.02 \e^{\frac{1 - u}{10}}$}
\label{variogram}
\end{center}
\end{figure}

We plot the resulting time dependent volatility (the so-called ``signature plot'') for an exponentially decaying $\mathcal{C}(u)$ in figure \ref{variogram}. One sees that positive correlations (trends) lead to an increase of the long-term volatility over the
short-term volatility, and vice-versa for negative correlations (mean reversion). The case of uncorrelated random walks leads, unsurprisingly, to a strictly constant volatility $\sigma(\tau)$. We will see in the
next section how trend following strategies in fact precisely capture the spread between long-term and short-term volatilities. 

\subsection{A toy model for the trend}
\label{toy_trend}

Let us consider a very simple trend strategy, which makes all the derivations very straightforward while keeping the main features of more sophisticated models. 
We consider a strategy such that the position $\Pi_t$ at time $t$ is proportional to the price difference between $t$ and $0$:
\be
\Pi_t := \lambda A_t \left ( S_t - S_{0} \right ) ,
\ee
where $\lambda$ is a certain factor which sets the risk level of the strategy and $A_t$ the amount of capital engaged, that we will set to $A_t \equiv 1$ henceforth.
The P\&L from $t-1$ to $t$ will then be (relatively to engaged capital):
\be
G_{t} := \Pi_{t-1} D_{t} = \lambda D_t \sum_{t'=1}^{t-1} D_{t'} \:\text{,}  \quad G_{1} := 0 \:\text{.}
\ee

Now we can aggregate the performance of our trending predictor from day $0$ to a given day $T$, and re-arrange the sums:
\begin{align}
\mathcal{G}_T := \sum_{t = 1}^{T} G_t &= \lambda \sum_{t = 2}^{T} D_t \sum_{t'= 1}^{t-1} D_{t'} = \lambda \sum_{\mathclap{\substack{ 1 < t \leq T \\ 1 \leq t' < t }}}  D_t D_{t'} \nonumber \\
                                      &= \frac{\lambda}{2} \left ( \sum_{t = 1} ^{T} D_t \right )^2 - \frac{\lambda}{2} \sum_{t = 1} ^{T} D_t^2 = \frac{\lambda}{2} \left ( {S_T - S_0} \right )^2 - \frac{\lambda}{2} \sum_{t = 1} ^{T} D_t^2
\label{TOY}
\end{align}

This simple formula is at the core of our understanding of the performance of a trending system.
In a nutshell, it says that the performance, aggregated over $T$ days, is proportional to the difference
between the realized variance computed over $T$ days and the variance computed using 1-day returns.
Using the definition of the above section, the {\it averaged} aggregated performance is therefore given by:
\be
\left\langle \sum_{t = 1}^{T} G_t \right\rangle = \frac{\lambda T}{2} \left ( \sigma^2(T) - \sigma^2(1) \right )  \:\text{,}
\ee
showing that, as expected from our discussion above, the performance of the trend is positive when the long-term volatility is larger than the short-term volatility, and vice-versa. 

These results calls for two further comments. First, the time unit is completely arbitrary, and is only defined by the frequency at which $\Pi_t$ is updated and the portfolio rebalanced. The resulting performance of 
the strategy is the difference between the long-term volatility and the volatility at the rebalancing time scale.

Second, we have defined $D_t$ as price differences. Following simple optimization procedures to manage the heteroskedasticity of financial time series, it is customary to normalize these price changes
by a local estimate of the volatility, such as price changes have constant variance (or as close to that as possible). A typical example is an exponential moving average (EMA) of the realized (daily) volatility over some time-scale $\tau_\sigma$: 
\be
\sigma_t := \gamma \sqrt {\mathcal{L}_{\tau_{\sigma}} \left [ D_t^2 \right ]}
\ee
with $\mathcal{L}_\tau [X_t] := (1 - \alpha) \sum_{i \leq t} \alpha^{t - i} X_i$ the EMA operator,
and $\alpha := 1 - {2}/({\tau + 1})$ which conventionally\footnote{see \url{https://en.wikipedia.org/wiki/Moving_average\#Exponential_moving_average}} defines $\tau$.
$\gamma$ is an empirical factor calibrated such that $R_t = D_t / \sigma_{t - 1}$ has unit variance.
In practice, we found $\gamma=1.05$ for $\tau_{\sigma}=10$, values we will keep throughout this paper.

All the above computations can be redone with normalized returns $R_t$ and lead to the performance of a {\it risk-managed trend strategy},
aggregated over $T$ days, which reads:
\be
\mathcal{G}_T := \sum_{t = 1}^{T} G_t = \frac{\lambda}{2}\left( \left(\sum_{t = 1} ^{T} R_t \right)^2 - \sum_{t = 1} ^{T} R_t^2 \right) \:\text{.}
\ee
Since by construction $\langle R_t^2 \rangle = 1$ (provided our volatility predictor is unbiased),
the last term of the equation is known on average.
This means that the performance of a risk-managed trend strategy is given by the normalized long-term variance of the underlying.

Consider a long history of this strategy, and split it in sub-samples of size $\tau$. The trend signal is reset to zero every $\tau$. 
Its average performance aggregated over a period $\tau$ can be plotted as a function of the (normalized) return of the underlying over the same time scale,
namely $\mathcal{T}_{t_i + \tau} := \sum_{t = 1} ^{\tau} R_{t_i + t} / \sqrt{\tau}$.
We have already pointed out that the second term in the above formula averages to 1, so we get the following conditional expectation of the performance:
\be\label{simple_trend}
\langle \mathcal{G} \given \mathcal{T} \rangle = \frac{\lambda \tau}{2} \, \left(\mathcal{T}^2 -1\right)  \:\text{.}
\ee
This formula is very simple, and clearly shows that we should expect positive convexity from such a simple strategy, but only over a certain time-scale: it should work well when the underlying asset has experienced large moves over the time horizon $\tau$ of the trend following strategy. On the other hand, it does {\it not} provide any protection against large moves taking place on time scales $\ll \tau$, such as overnight market gaps for example. 

\subsection{Trend following using EMAs}

In practice, most trend following funds use combination of EMAs to compute their trending signals \cite{DG_JR}. Therefore, it is important to generalize and extend our formula above to these types of filters. 
In this subsection, we consider a very standard trend-following strategy, which uses a single EMA of past returns $\mathcal{L}_{\tau}[R_t]$ to determine the position $\Pi_t$ as:
\be \label{position}
\Pi_t := \frac{\lambda \tau \mathcal{L}_{\tau}[R_t]}{\sigma_t}.
\ee

Defining again $G_{t} := \Pi_{t-1} D_t$ as the instantaneous gain, one obtains the following exact result for a certain EMA of this gain (see Appendix \ref{apdx:discrete_trend} for a proof):
\begin{equation}
\mathcal{L}_{\tau'} [G_{t}] = \frac{\lambda \tau}{\tau-1} \left(\tau \mathcal{L}^2_{\tau}[R_t]  - \mathcal{L}_{\tau'} \left [R_t^2 \right]\right )
\label{theorem_ema}
\end{equation}
with $\tau' = \frac{\tau}{2} + \frac{1}{2\tau} \approx \tau/2$, where the last approximation is valid when $\tau \gg 1$.
This equation is, to the best of our knowledge, new and tells us that the trend-following performance,
once averaged over a suitable period of time (roughly half of the trend following strategy itself),
can be rewritten exactly as a difference between a long term volatility (the square of the EMA of past returns) and a short term volatility
(the EMA of the square of daily returns). Note however that the first EMA is with a time scale $\tau$, whereas the second one is on scale $\tau'$.

\begin{figure}
\begin{center}
\epsfig{file=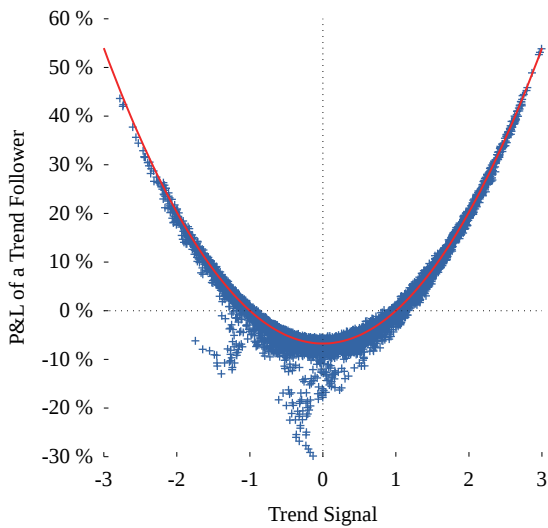}
\caption{Aggregated P\&L of trend following $\mathcal{G}$ as a function of $\mathcal{T}$
(i. e. the past trend computed as an EMA) for S\&P 500 futures, rolled from Januray 1983 to October 2015.
Here we use $\tau=180$ and $\lambda = 0.01 / \sqrt{\tau}$ such as the daily P\&L has a standard deviation of 1\%.
}
\label{SPema}
\end{center}
\end{figure}

Rather than the average daily performance, it is useful to consider
\be\nonumber
\mathcal{G}_{t} := \tau' \mathcal{L}_{\tau'} [G_{t}],
\ee
which is closer to the aggregated performance over a period $\tau'$.
If our volatility estimate is un-biased and ensures that the second term remains close to unity, we finally find:
\begin{equation}
\left \langle \mathcal{G} \given \mathcal{T} \right \rangle = \Upsilon(\tau) \left(\mathcal{T}^2 - 1 \right)
\label{EMAconv}
\end{equation}
with $\mathcal{T} := \sqrt{\tau} \mathcal{L}_{\tau}[R_t]$ and $\Upsilon(\tau) := \frac{\lambda \tau \tau'}{\tau-1} $.
Note that when $\tau \gg 1$, $\Upsilon(\tau) \approx \frac{\lambda \tau}{2}$,
such that equation \ref{simple_trend} is, {\it mutatis mutandis}, recovered.
As an illustration of this formula, we have plotted in figure \ref{SPema} the aggregated performance over $\tau'$ days, $\mathcal{G}$,
of a trend strategy defined by $\tau=180$ (corresponding to $\tau' \approx 90$).
As we can see, we get a very good agreement between the theoretical line and the realized performance.
In particular, the convexity is quite visible in this set-up, while it would have been utterly blurred had we looked at 1-day, or even 1-month, returns.
This sensitivity to the averaging time-scale means that we need to estimate carefully the horizon CTAs use to define their trend if we want to detect any sign of convexity in the CTA performance.
This is what we will attempt to do in section \ref{reconstructing}.

\subsection{Capping the position}
\label{capping_the_position}

\begin{figure}
\begin{center}
\epsfig{file=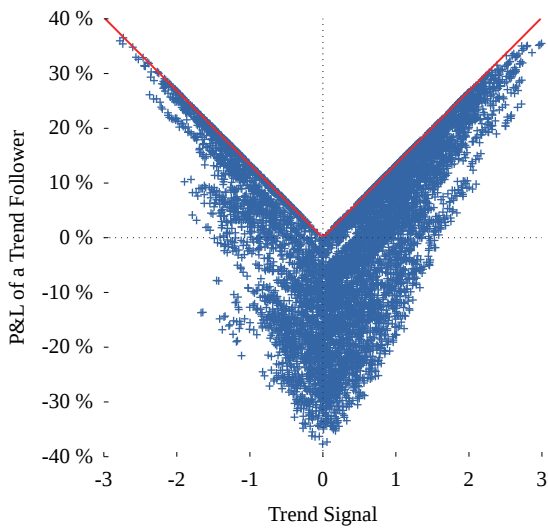}
\caption{$\mathcal{G}$ as a function of $\mathcal{T}$
when the position is equal to the
sign of $\mathcal{T}$ (same conditions as in figure \ref{SPema}).
The predicted V-shape is very clearly observed.
}
\label{TRENDsign}
\end{center}
\end{figure}

Practitioners in fact use a variety of different signals to capture the trend anomaly: square average instead of exponential, Vertical-Horizontal Filters, strength indices, crossing of moving price averages, 
non-linear transformation of EMAs, etc. We want to present here a case of particular interest: what happens if the position is simply $\Pi_t = \pm 1$ depending on the sign of the past trend, rather than proportional to it?
The idea of {\it capping} the position is quite common in practice, since it avoids taking large positions that can lead to uncontrolled losses. The $\pm 1$ solution can be viewed as extreme capping procedure.  
One can show that in that case, the performance of the trend-following strategy can be written in the following form (see \ref{apdx:non_linear_trend}):
\begin{equation}
\left \langle \mathcal{G} \given \mathcal{T} \right \rangle = \lambda \tau \left ( |\mathcal{T}| - \sqrt{\frac{2}{\pi}} \right) \:\text{.}
\label{SIGNconv}
\end{equation}
We have plotted in figure \ref{TRENDsign} the typical shape of the aggregated performance over $\tau$ days, $\mathcal{G}$,
as a function of the trend indicator $\mathcal{T}$.
Here as well, the performance shows a convex shape as a function of the trend indicator.
Instead of a parabolic fit, however, we see a piece-wise linear profile.

For less extreme capping procedures (for example $\Pi_t \propto \tanh(\mathcal{T}_t)$)
one can argue that the shape of $\mathcal{G}$ vs. $\mathcal{T}$ interpolates between the parabola of equation \ref{simple_trend},
valid for small $\mathcal{T}$, and the V-shaped curve above, valid for large $\mathcal{T}$.
In any case, the positive convexity of this curve is a generic property of trend following strategies.

\subsection{Skewness of trend following strategies}

The skewness of the profit and loss distribution of a given strategy was recently proposed as a natural discriminant between risk premia strategies and genuine market anomalies (see \cite{YL_RP} for a thorough discussion). 
Several previous papers have pointed out that the skewness of trend following is positive (see \cite{JP_MP, RM_DZ}, and \cite{BB_NG} for a general framework), at variance with risk premia that have a negative skewness. 
Using the above results, we see that the positive skewness follows directly from Eqs. (\ref{simple_trend}) and (\ref{EMAconv}), at least on the time scale of the trend following distribution. Indeed, one can assume that $\mathcal{T}$ is
approximately Gaussian (which is justified by the central limit theorem when $\tau$ is large) which immediately tells us that the performance of trend following on scale $\tau$ has a $\chi^2$ distribution, which has a known positive skewness. In fact, this result is general and does not even depend on the fact that trend following leads to positive gains on average. If the returns are completely uncorrelated, clearly trend following is a zero gain strategy, but with positive skewness. In other words, trend followers lose more often than they gain, as noted in \cite{JP_MP}.

The skewness of the daily returns of a trend following strategy on scale $\tau$ is more subtle. Assuming that the covariance of daily returns, defined in equation (\ref{correl}), is given by $\mathcal{C}(u)=q^u$ with $0 < q \ll 1$, one finds that
the skewness of the daily returns is $6q + O(q^2)$, i.e. it is positive if returns are indeed auto-correlated in time, but zero for a pure random walk ($q=0$) as was already noticed in \cite{JP_MP}.

\subsection{Summary}

We have seen that the performance of a trend following strategy, once aggregated over a suitable time horizon, can be thought of as the difference between a long-term and a short-term variance. If properly risk-managed, the short-term variance is on average constant, and therefore the P\&L is directly related to the square of the past long-term return. 
In other words, this single-asset strategy has a strong positive convexity and therefore a positive skewness on the long term.

Let us emphasize again that this conclusion does not depend on the overall performance of the trend itself. One can assume returns to be perfectly uncorrelated, and hence the trend to have vanishing expected returns -- still, there is positive convexity and positive skewness once the performance is aggregated over a proper time scale. This conclusion holds independently of the precise implementation of the trend strategy.

It is also interesting to point out the importance of the re-balancing period of the strategy. For a $\tau$-day trend, implemented every $n$ days, one can expect some tail protection when the market moves a lot over $\tau$ days, but
this protection disappears for large sudden moves happening on a time smaller than $n$ days -- the strategy is just blind to these high frequency oscillations.

\section{Convexity at work on CTA indices}

In this section, we want to understand how the above analysis can be used to scrutinize the performance of the  CTA industry. We will consider the SG CTA Index, and show that our simple trend strategy allows us to reproduce its main features (as previously shown in \cite{NB_RK}), provided an appropriate value of the effective
time horizon $\tau$ is chosen. We show that the convexity measured on this time scale is much larger than that shown in the introduction (see figure \ref{BTOPconv}-Right), but not as high as predicted in the previous section (figure \ref{SPema}). This can be traced back to the fact that CTAs do not operate on a single contract, but apply trend strategies to a large pool of contracts, with different degrees of correlations.

\subsection{Reconstructing the SG CTA Index}
\label{reconstructing}

We first want to compare the performance of a simple trend following system to that of the best known CTA index: the SG CTA Index.\footnote{Previously known as the NewEdge CTA Index.}
For simplicity, we only included our simulations futures contracts corresponding stock indices, government bonds, short term interest rates, foreign exchange rates, and commodities. For each of these asset classes, we considered the most liquid contracts. We end up with a selection outlined in table \ref{FutNEI}. We believe that this selection is un-biased, since liquidity is quite stable over time, and these contracts have been available for at least a few decades. In any case, adding or removing a few contracts should not significantly affect our results.

\begin{figure}
\begin{center}
\begin{tabular}{|c|c|c|}
\hline
Commodities           & Stock Indices        & Foreign Exchange Rates \\
\hline
WTI Crude Oil (CME)   & S\&P 500 (CME)       & EUR/USD (CME)          \\
Gold (CME)            & EuroStoxx 50 (Eurex) & JPY/USD (CME)          \\
Copper (CME)          & FTSE 100 (ICE)       & GBP/USD (CME)          \\
Soybean (CME)         & Nikkei 225 (JPX)     & AUD/USD (CME)          \\
                      &                      & CHF/USD (CME)          \\
\hline
\end{tabular}
\begin{tabular}{|c|c|}
\hline
Short term interest rates    & Government bonds                                     \\
\hline
Euribor (ICE)                & 10Y U.S. Treasury Note (CME)                         \\
Eurodollar (CME )            & Bund (Eurex)                                         \\
Short Sterling (ICE)         & Long Gilts (ICE)                                     \\
                             & JGB (JPX)                                            \\
\hline
\end{tabular}
\caption{Most liquid futures in each sector, included in our simulation.
This selection is stable across time.
The time series are considered between January 2000 and October 2015.
}
\label{FutNEI}
\end{center}
\end{figure}

The position of our surrogate CTA is asset $k := 1,\dots, N$ at time $t$ is the natural generalisation of Eq. (\ref{position}) above:
\begin{equation}
\Pi_{k,t} := \lambda A_t \tau \frac{w_k \mathcal{L}_{\tau}[R_{k,t}]}{\sigma_{k,t}} \:\text{,}
\end{equation}
where $\lambda$ is a certain factor which sets the risk level of the diversified strategy,
$w_k$ the weight of asset $k$ in the portfolio, and we again set  the amount of capital engaged equal to unity ($A_t=1$),
and $\mathcal{L}_{\tau}[R_{k,t}]$ is the EMA over the returns of asset $k$
(normalized by their local volatility) over time $\tau$.
The total gain of the portfolio then reads:

\begin{equation}
\mathbf{G}_t := \sum_{k} G_{k,t} = \sum_k {\Pi_{k,t-1} D_{k,t}} = \lambda \tau \sum_k w_k \mathcal{L}_{\tau}[R_{k,t-1}] R_{k,t} \:\text{.}
\label{Geq}
\end{equation}

To compare the performance $\mathbf{G}_t$ of our surrogate CTA with that of the SG CTA Index (called $\mathbf{G}_t^{\text{CTA}}$),
we need to determine the value of $\lambda$ and $\tau$, and of the different weights $w_k$.
For the sake of simplicity, we do not optimize these weights to maximize the correlation between $\mathbf{G}_t^{\text{CTA}}$ and $\mathbf{G}_t$,
but allocate the same amount of risk on each asset: $w_k = 1/N$.

\begin{figure}
\begin{center}
\epsfig{file=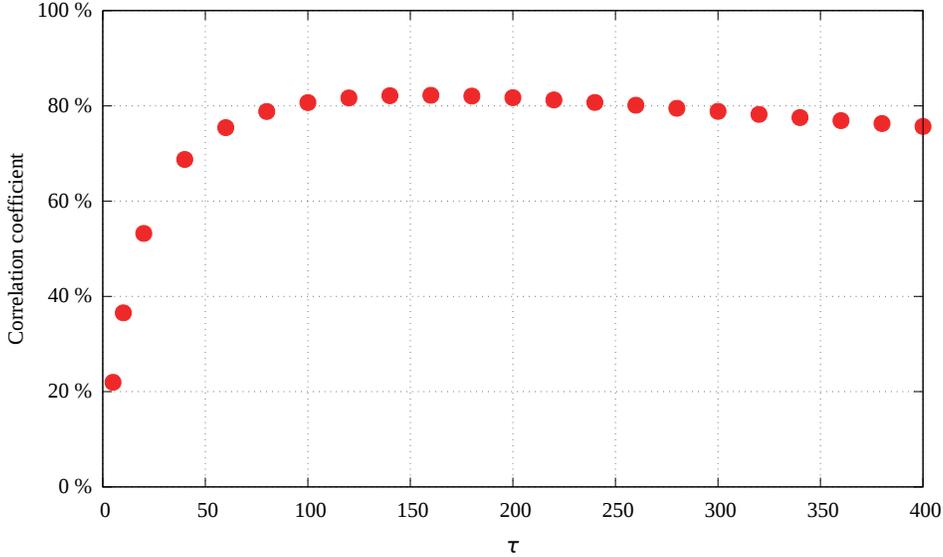}
\caption{correlation between $\widetilde{\mathbf{G}}_t$ and the SG CTA Index as a function of the time-scale of the trend we used $\tau$.
As we can see, the maximum is around $\tau=180$ days.
}
\label{corrNEI}
\end{center}
\end{figure}

To be fully accurate, we also need to make some assumptions on the fee structure.
We assume flat transaction costs $c_t = 2\%$, and a typical 1\% management fee-20\% incentive fee $f_t$.
We also consider the S\&P/BGCantor 3-6 Month U.S. Treasury Bill Index\footnote{Ticker: SPBDUB6T} as the risk-free rate $r_t$,
and define $\widetilde{\mathbf{G}}_t := \mathbf{G}_t - c_t - f_t + r_t$ as the total net performance of our surrogate fund.
The value of $\lambda$ is then obtained such that the volatility of $\mathbf{G}_t^{\text{CTA}}$ and $\widetilde{\mathbf{G}}_t$ are equal.
To determine the relevant time scale $\tau$, we maximize the correlation between $\mathbf{G}_t^{\text{CTA}}$ and $\widetilde{\mathbf{G}}_t$.
The value of the correlation as a function of $\tau$ is shown in figure \ref{corrNEI}.
As we can see, there is a broad maximum around $\tau=180$.
We choose this value as the typical time scale of the CTA trend following strategies.

\begin{figure}
\begin{center}
\epsfig{file=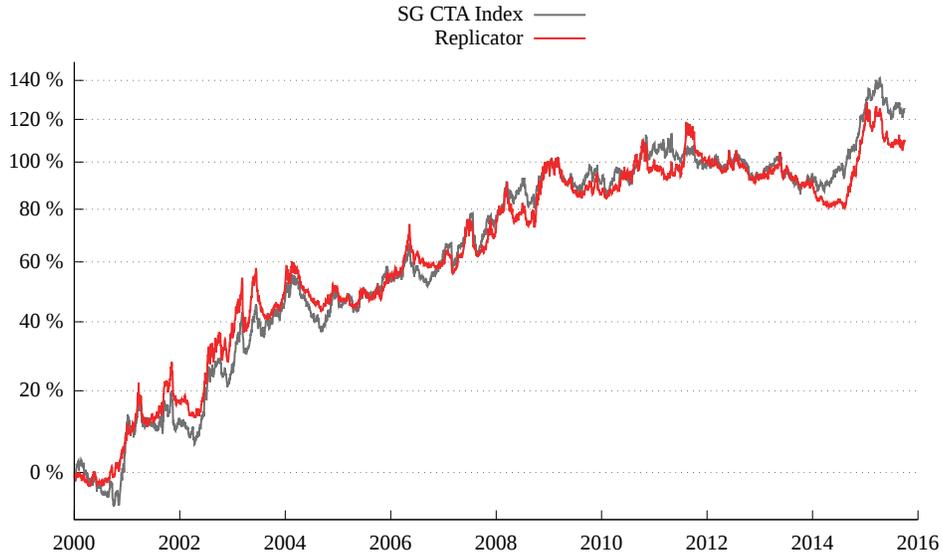}
\caption{Cumulated returns of $\widetilde{\mathbf{G}}_t$ and the SG CTA Index.
We seem to capture all the alpha contained in the SG CTA Index with our simple replicator.
}
\label{NEIpnl}
\end{center}
\end{figure}

What is striking is the high value of the resulting correlation (above 80\%),
when all we did was to follow a basic, un-sophisticated trend signal on the most liquid assets on the planet.
Figure \ref{NEIpnl} shows that we actually capture most of the alpha contained in the SG CTA Index,
since its Sharpe ratio is very close to that of our surrogate CTA.
A still higher correlation could be achieved with a more sophisticated risk allocation (like in \cite{CL_al} for example),
but this is beyond the scope and purpose of the present work.

\subsection{The convexity of the SG CTA Index}

\begin{figure}
\begin{center}
\epsfig{file=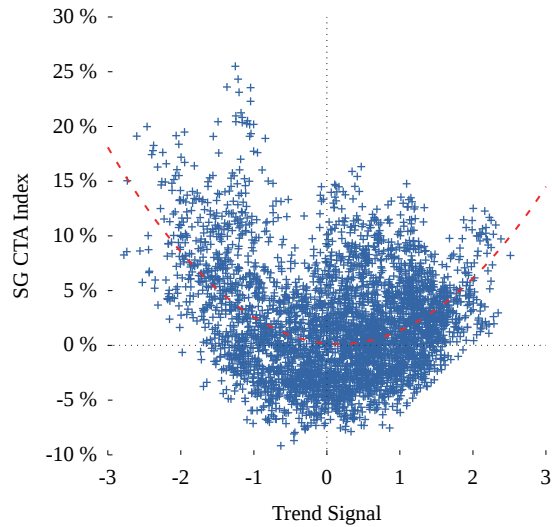}
\caption{Aggregated performance over $\tau'$ days of the SG CTA Index as a function of a trend on  S\&P 500 futures,
$\mathcal{T}_{\text{SP}} := \sqrt{\tau} \mathcal{L}_{\tau}[R_{\text{SP}, t}]$,
with $\tau=180$ and $\tau' \approx 90$.
$R^2=0.18$, which is significantly higher than reported in figure \ref{BTOPconv}-Right.}
\label{NEIconvexity}
\end{center}
\end{figure}

Now we have determined the time horizon over which CTA operate, we can directly test our prediction, Eq. (\ref{EMAconv}),
on the convexity of trend following -- at least with respect to the equity market, here the S\&P 500 Index.
In other words, we want to plot the aggregated CTA performance $\bm{\mathcal{G}}^{\text{CTA}}_t := \tau' \mathcal{L}_{\tau'} [\mathbf{G}_t^{\text{CTA}}]$
(over time $\tau' \approx 90$ days) as a function of a trend on the return of the S\&P Index
over time $\tau \approx 180$ days, $\mathcal{T}_{\text{SP}, t} := \sqrt{\tau} \mathcal{L}_{\tau}[R_{\text{SP}, t}]$.
The result is plotted in figure \ref{NEIconvexity}. As we can see, the convexity is much more pronounced than in figure \ref {BTOPconv}-Right. In particular, the $R^2$ of the quadratic fit suggested by equation \ref{EMAconv} increases from a meager $0.02$ to a more convincing value $\approx 0.2$.
So there is some convexity in the SG CTA Index, provided
appropriate averaging time scales are chosen. The still relatively low value of the $R^2$ can intuitively be understood: it comes from the fact trend following is done on a variety of contracts, while there is an unique reference 
contract (here S\&P 500 futures). We therefore need to understand how diversification affects the convexity of the performance of CTAs.

\subsection{Convexity and diversification}

The natural idea is not to take S\&P 500 Index as the reference, but rather a long-only, risk managed diversified portfolio,
and show that a diversified trend following strategy is convex with respect to this product.
Namely, we consider positions $\propto 1/\sigma_{k,t-1}$ and weights $\omega_k$ on asset $k$,
so that the returns of this portfolio read:
\begin{equation}
\mathbf{G}_t^{\text{RP}} := \sum_k \omega_k R_{k,t}
\label{Grp}
\end{equation}
This portfolio is long everything we have in our universe: indices, bonds, commodities, and a basket of currencies against the dollar.\footnote{The currency basket is in fact not needed 
but simplifies the mathematical analysis.} Again for simplicity, we choose $\omega_k = w_k = 1/N$, such that the risk taken on every asset is the same. Since this portfolio achieves the same risk on each of its components, 
is has become customay to call it a ``Risk Parity'' portfolio \cite{TR_RP}.  Incidentally, we find that the above portfolio replicates surprisingly well standard Risk Parity indices, 
such as the J.P. Morgan Index\footnote{JPMorgan's index of 17 risk parity funds that report daily, data from JP Morgan.} with which it has an 89\% correlation.

We now want to show that the diversified trend CTAs provides a hedge for this portfolio,
in the sense that a strict bound on its convexity can be established.
To do so, we introduce a trend on the P\&L of our Risk Parity portfolio on scale $\tau$,
\begin{equation}
\bm{\mathcal{T}}^{\text{RP}}_t := \sqrt{\tau} \mathcal{L}_{\tau}[\mathbf{G}_t^{\text{RP}}] = \sum_k w_k \mathcal{T}_{k,t} \:\text{,}
\end{equation}
and find using a simple convexity argument that the aggregated P\&L of our CTA replicator $\bm{\mathcal{G}}$ is bounded from below:
\begin{align}
\bm{\mathcal{G}}_t &:= \sum_{k} w_k \mathcal{G}_{k,t} \\
                   &=   \Upsilon (\tau) \sum_{k} w_k \left ( \tau \mathcal{L}_{\tau}^2[R_{k,t}] - \mathcal{L}_{\tau'}[R_{k,t}^2] \right ) \\
                   &\ge \Upsilon (\tau) \left ( \tau \mathcal{L}_{\tau}^2 \left [ \sum_{k} w_k R_{k,t} \right ] - \sum_{k} w_k \mathcal{L}_{\tau'}[R_{k,t}^2] \right ) \\
                   &\ge \Upsilon (\tau) \left ( \left(\bm{\mathcal{T}}^{\text{RP}}_t\right)^2 - \sum_{k} w_k \mathcal{L}_{\tau'}[R_{k,t}^2] \right )
\end{align}
Finally leading us to:
\begin{equation}
\left \langle \bm{\mathcal{G}} \given \bm{\mathcal{T}}^{\text{RP}} \right \rangle \ge \Upsilon(\tau) \left( \left(\bm{\mathcal{T}}^{\text{RP}}\right)^2-1 \right) \:\text{.}
\label{GGRP}
\end{equation}

\begin{figure}
\begin{center}
\epsfig{file=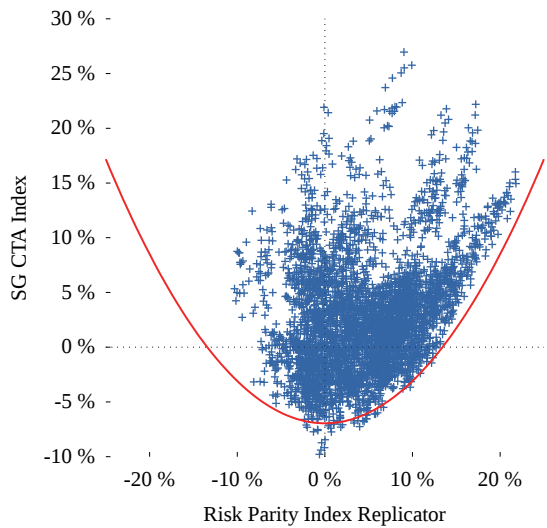}
\caption{Aggregated CTA performance $\bm{\mathcal{G}}^{\text{CTA}}$ as a function of the aggregated Risk-Parity performance $\sqrt{\tau} \bm{\mathcal{T}}^{\text{RP}}$.
We use our proxies for both performances, using the pool of contracts listed in table \ref{FutNEI}.
We observe that all points lie above the parabola predicted by the inequality (\ref{GGRP}) (in red).
}
\label{RPrep2}
\end{center}
\end{figure}

Note that the right hand side is giving exactly the expected average performance of a trend following strategy on the ``Risk Parity'' portfolio -- see figure \ref{RPrep2}.
This means that CTAs do provide a very significant protection against the potential large moves of our proxy Risk Parity portfolio
which, as mentionned above, is highly correlated with the J.P. Morgan Risk Parity Index benchmark.

\section{Trend following and option strategies}
We now turn to another feature of trend following strategies: its ``long-option'' like feature emphasized in numerous papers.
This is a very natural connection, since both strategies are characterized by a positive convexity and a positive skewness.
We want to discuss this analogy more precisely within our framework above.
We will in particular propose an option portfolio that captures long-term volatility in a model-free way,
completely independent of any assumption on the underlying, in particular not relying on any Black-Scholes hocus-pocus.
The pay-off of that portfolio is actually identical to that of the simple trend following strategy discussed above.
We will see how the hedge of this portfolio essentially exchanges the long-term variance for the short-term one.

\subsection{A collection of strangles}
\label{collection_strangles}
We begin our study by considering a simple straddle, composed of an At The Money (ATM) Put option and an ATM Call option, both of maturity $T$ (similar to what was considered in \cite{IM}).
Here and throughout this paper, we neglect all interest rate effects and therefore set the interest rate to zero.
It is quite straightforward to verify that the P\&L of this portfolio can be written as:
\be
\mathcal{G}_T^{\text{straddle}} := |S_T-S_0| - (C_{0,T}^{S_0} + P_{0,T}^{S_0}) \:\text{,}
\ee
where $S_0$ is the current spot price (equal to the strike $K$ of both the Call and the Put) and $C_{t,T}^{K}, P_{t,T}^{K}$ are respectively the price paid for the Call and the Put at time $t$.
This pay-off is very similar to that of the capped trend following strategy described in section \ref{capping_the_position}. To obtain an expression closer to Eq. (\ref{TOY}), we need to consider an infinite, equi-weighted collection of strangles of strike price $K$, centered at the money, such that the pay-off of the portfolio is:
\begin{align*}
\underbrace{\int_0^{S_0} (K - S_T)_{+} dK}_{= \frac{1}{2} \left [ (S_0 - S_T)_{+} \right ]^2} + \underbrace{\int_{S_0}^\infty (S_T-K)_{+}dK}_{= \frac{1}{2} \left [(S_T - S_0)_{+} \right ]^2} \equiv \frac{1}{2} (S_T - S_0)^2
\end{align*}
The total profit associated with this portfolio can thus be written as
\begin{align}
\mathcal{G}_T^{\text{strangles}} &:= \frac{1}{2} (S_T - S_0)^2 - \int_0^{S_0} P_{0,T}^K + \int_{S_0}^\infty C_{0,T}^K \nonumber \\
                                 &= \frac{1}{2} \left ((S_T - S_0)^2 - T\bar{\sigma}_{0,T}^2\right ) \:\text{,}
\label{STRpo}
\end{align}
where $\bar{\sigma}_{0,T}$ is an effective implied volatility defined by the second equality. Equation (\ref{STRpo}) is independent of the assumed model for the dynamics of the underlying and is quite suggestive. It tells us that this portfolio receives the long-term variance $(S_T - S_0)^2$, and pays a fixed price at the start of the trade set by the implied volatility level. Note that the long-term variance is exactly the same as that of the toy trend-following model, Eq. (\ref{TOY}). The only difference between the two strategies is the price
paid to receive this variance: short-term realized volatility in the latter case vs. implied volatility for the strangle portfolio. Since options are notoriously sold at a premium, the option-based strategy would be more expensive than the trend-following one.

In practice, only a finite set of strikes are available, so the above theoretical portfolio cannot be exactly reproduced. However, just like any capped trend-following strategy interpolates between the V-shaped and the parabolic pay-off (see Eqs. \ref{TOY} and \ref{TRENDsign}), the pay-off of an approximately uniform collection of strangles should lie somewhere in between the pay-off the above infinite collection of strangles and that of a single straddle. 
The quadratic pay-off crosses over to an asymptotically linear pay-off takes beyond the maximum and minimum strikes available in the market. 

\subsection{To hedge is to trend}

\begin{figure}
\begin{center}
\epsfig{file=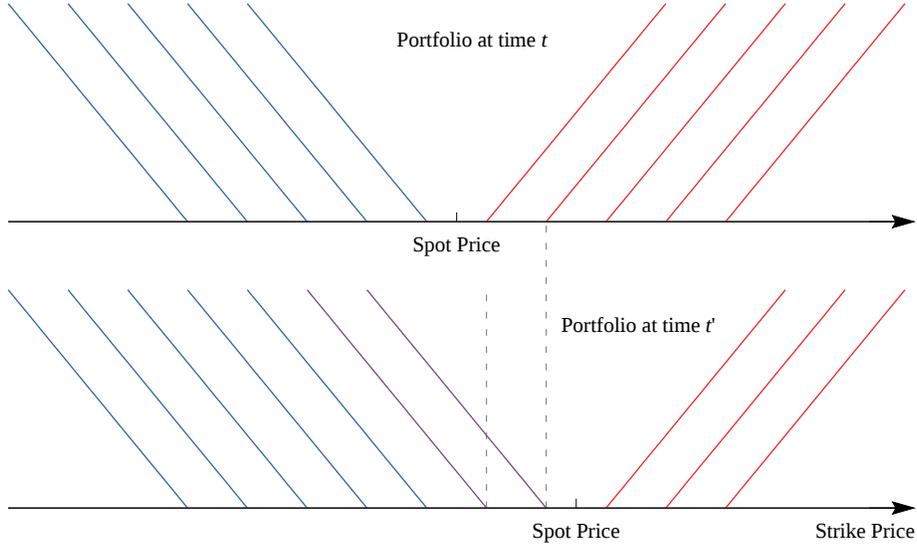}
\caption{Re-balancing uniform-weight portfolio by trading futures between two spot prices}
\label{fig: strangles_rebalance}
\end{center}
\end{figure}

At $t=0$, the infinite strangle portfolio is approximately Delta neutral,\footnote{It is exactly Delta neutral within a symmetric additive model with zero interest rates.} since all strangles centered at the money.
One can try to intuitively understand the hedging trade needed to remain Delta neutral when the spot price moves from $S_0$ to $S_t$.
Figuratively, we have to re-center our portfolio of strangles around the new price $S_t$, so we have to exchange Calls for Puts (resp. Puts for Calls) if the price has gone up (resp. down) between time $0$ and $t$.
But selling a Call and buying a Put at the same strike is strictly selling the underlying, so the required hedging is sell the underlying by an amount $S_t - S_0$ when the price goes up,
and to buy it if it goes down (see figure \ref{fig: strangles_rebalance} for an illustration).

As the alert reader will have realized, this hedging strategy is exactly the mirror image of the toy-model for trend-following considered in section \ref{toy_trend}. Hence, the
total performance of the hedge from $t=0$ to the maturity of the options is given by:
\be
\mathcal{G}_T^{\text{hedge}} := \frac12 \sum_{t=1}^T D_t^2 - \frac12  (S_T - S_0)^2
\ee
This means that the hedge exchanges the variance of the return over the period $[0, T]$ with the variance defined using the re-hedging frequency (see \cite{BD} for a detailed account on the role of hedging, and \cite{BP_OPT} for a discussion of the impact of the hedging frequency when returns are correlated).

If we add the above pay-off of our infinite strangle portfolio to this hedge P\&L,
we find that the total performance of the hedged portfolio of options is:
\begin{equation}
\label{VS}
\mathcal{G}_T^{\text{strangles}} + \mathcal{G}_T^{\text{hedge}} = \frac12 \sum_{t=1}^T D_t^2 - \frac{T}{2}  \bar{\sigma}_{0,T}^2
\end{equation}
This shows that the performance of our hedged strangle portfolio is very similar to the one of a standard variance swap,
but note that equation (\ref{VS}) is exact and model independent, whereas the standard $1/K^2$ variance swap portfolio relies of a Black-Scholes formula (see for example \cite{ED, PC} for detailed reviews on the topic).
As originally discussed in \cite{BP_OPT} in this context,
it is the hedging frequency that determines which volatility we are exposed to.
If one re-hedges every 5 minutes, the performance will be determined by the realized volatility of 5-minutes returns,
while one will capture the realized daily volatility if the re-hedge is done at the close of each day.
The price paid for the options, on the other hand, is always the same and determined by the implied volatility $\bar{\sigma}_{0,T}$ at time $t=0$.

\subsection{Summary}

The above results help us clarify the similarities and differences between a trend following system and a long naked strangle portfolio.
Both give exposure to long term variance (i.e. they both provide protection for a long-stock portfolio),
but the second one buys that exposure at a fixed price $\bar{\sigma}_{0,T}$ while the first one pays the short-term variance $\sum_{t=1}^T D_t^2$.
An interesting consequence is that, if a large shock occurs in one single day that dominates the total return over the interval $[0, T]$,
the naked straddle will make money, since the entry price is fixed, while the trend strategy will be flat on average.
In other words, the straddle is a better hedge, and therefore its price $\bar{\sigma}_{0,T}$ should be higher than the realized volatility $\sum_{t=1}^T D_t^2$.

The premium paid on option markets is however too high in the sense that long-vol portfolios have consistently lost money over the past 2 decades (barring the 2008 crisis),
while trend following strategies have actually posted positive performance. So, even if options provide a better hedge, trend following is a much cheaper way to hedge long-only exposures.

\section{Perspectives}

In this paper, we have shown that single-asset trend strategies have built-in convexity provided its returns are aggregated over the right time-scale, i.e, that of the trend filter. In fact, the performance of trend-following can be viewed as a swap between a long-term realized variance (typically the time scale of the trending filter) and a short-term realized variance (the rebalancing of our portfolio). This feature is a generic property and holds for various filters and saturation levels. While trend following strategies provide a hedge against large moves unfolding over the long time scale, it is wrong to expect a 6 to 9 months trending system rebalanced every week to hedge against a market crash that lasts a few days. 

We dissected the performance of the SG CTA Index in terms of a simple replication index, using an un-saturated trend on a equi-weighted pool of liquid assets. Assuming realistic fees, and fitting only the time-scale of the filter (found to be of the order of 6 months) we reached a very strong correlation (above 80\%) with the SG Index, and furthermore fully captured the average drift (i.e. our replication has the same Sharpe ratio as the whole of the CTA industry). However, our analysis makes clear that CTAs do not provide the same hedge single-asset trends provide: some of the convexity is lost because of diversification. We however have found that CTAs do offer an interesting hedge to Risk-Parity portfolios. This property is quite interesting, and we feel it makes the trend a valid addition in the book of any manager holding Risk Parity products (or simply a diversified long position in both equities and bonds). 

Finally, we turned our attention to the much discussed link between trend-following and long-volatility strategies. We found that a simple trend model has exactly the same exposure to the long-term variance as a portfolio of naked strangles. The difference is the fact that the entry price of the latter is fixed by the implied volatility, while the cost of the trend is the realized short-term variance. The pay-off of our strangle portfolio is {\it model-independent} and coincides with that of a traditional variance swap -- except that the latter requires Black-Scholes-like assumptions.

All in all, our results prove that trending systems offer cheap protection to long-term large moves of the market. This, coupled with the high statistical significance of this market anomaly \cite{YL_TR,AQR,WIN}, really sets trend following apart in the world of investment strategies. A potential issue might be the global capacity of this strategy, but recent performance seems to be quite in line with long-term returns, so there is at present little evidence of over-crowding.

\section{Appendix}

\subsection{Proof of trend formula}
\subsubsection{Theorem}

Let $X_t$ be a discrete process and $\mathcal{F}_\alpha$ be the following linear filter:
\begin{equation}
\label{operator}
\mathcal F_\alpha [X_t] = \sum_{i\geq 0} \alpha^i X_{t-i} \quad (\alpha >0).
\end{equation} 
we have the following relation:
\begin{equation}
 \mathcal F_{\alpha\beta} \Big[ Y_t\mathcal F_\alpha [X_t] +  X_t\mathcal F_\beta[Y_t]\Big]  = \mathcal F_\alpha [X_t] \mathcal F_\beta [Y_t]   +\mathcal F_{\alpha\beta} [X_tY_t].
\end{equation}

\textbf{Proof}\medskip

Using the definition of the filter, we decompose the first term of the left hand-side:
\begin{align}\label{drift}
\mathcal F_\alpha [X_t] \mathcal F_\beta [Y_t] = \sum_{i\geq 0} \alpha^i\beta^i  X_{t-i}Y_{t-i} &+ \sum_{i\geq 0} \sum_{j>i} \alpha^i\beta^j X_{t-i}Y_{t-j} \\
                                             &+ \sum_{j\geq 0} \sum_{i>j} \alpha^i\beta^j X_{t-i}Y_{t-j}.
\end{align}

The cross term:
\begin{align}
\label{cross1}
\sum_{i\geq 0} \sum_{j>i} \alpha^i\beta^j X_{t-i}Y_{t-j}
  &= \sum_{i\geq 0}\alpha^i\beta^iX_{t-i} \sum_{j>i} \beta^{j-i}Y_{t-i -(j-i)} \nonumber \\
  &= \sum_{i\geq 0}\alpha^{i}\beta^iX_{t-i} \sum_{k\geq 1} \beta^{k}Y_{t-i-k} \\
\text{with } k = j-i \nonumber \\
  &= \sum_{i\geq 0}  \alpha^{i}\beta^i \left( X_{t-i} \mathcal F_\beta [Y_{t-i}] - Y_{t-i}\right)  \nonumber \\
  &= \sum_{i\geq 0}  \alpha^{i}\beta^i X_{t-i}\mathcal F_\beta [Y_{t-i}]  - \sum_{i\geq 0}  \alpha^{i}\beta^i X_{t-i}Y_{t-i} \nonumber \\
  &= \sum_{i\geq 0}  (\alpha\beta)^{i} U_{t-i} - \sum_{i\geq 0}  (\alpha\beta)^{i} X_{t-i}Y_{t-i} \\
\text{with } U_{t} = X_t\mathcal F_\beta [Y_t] \nonumber \\
  &= \mathcal F_{\alpha\beta} [U_t] -\mathcal F_{\alpha\beta} [X_tY_t] \\
  &= \mathcal F_{\alpha\beta} \Big[X_t\mathcal F_\beta [Y_t]\Big] -\mathcal F_{\alpha\beta} [X_tY_t].
\end{align}
Apply the same trick to the other cross term, we have:
\begin{equation}\label{cross2}
\sum_{j\geq 0} \sum_{i>j} \alpha^i\beta^j X_{t-i}Y_{t-j} =\mathcal F_{\alpha\beta} \Big[Y_t\mathcal F_\alpha [X_t]\Big] -\mathcal F_{\alpha\beta} [X_tY_t].
\end{equation}

The diagonal term:
\begin{equation}
\label{diagonal}
\sum_{i\geq 0} \alpha^{i}\beta^i X_{t-i}Y_{t-i} =\mathcal F_{\alpha\beta} [X_tY_t].
\end{equation}
\noindent
Using (\ref{drift}), (\ref{cross1}), (\ref{cross2})and (\ref{diagonal}), we finally have:
\begin{equation}
\label{filter1}
\mathcal F_{\alpha\beta} \Big[Y_t\mathcal F_\alpha [X_t] +  X_t\mathcal F_\beta[Y_t]\Big]  = \mathcal F_\alpha [X_t] \mathcal F_\beta [Y_t]   +\mathcal F_{\alpha\beta} [X_tY_t].
\end{equation}
We remark that $\mathcal F_\alpha [X_t]  = X_t + \alpha\mathcal F_\alpha [X_{t-1}] $, we can rewrite the above formula as:
\begin{equation}
\label{filter2}
\mathcal F_{\alpha\beta} \Big[\alpha Y_t\mathcal F_\alpha [X_{t-1}] +  \beta X_t\mathcal F_\beta[Y_{t-1}]\Big]  = \mathcal F_\alpha [X_t] \mathcal F_\beta [Y_t]   - \mathcal F_{\alpha\beta} [X_tY_t].
\end{equation}

\noindent

\subsubsection{Discrete equation for EMA filter}\label{apdx:discrete_trend}
We now employ the above result to derive the main result given in Section 2 for an exponential moving average filter. Taking $\alpha = 1- 2/(\tau + 1)$, and renormalize the filter $\mathcal{F}_\alpha$ in a way that the sum of weights is equal to 1. Let $\mathcal{L}_\tau$ be the renormalized filter:
\begin{equation}
\mathcal{L}_\tau \equiv (1 - \alpha)\mathcal{F}_\alpha.
\end{equation}
The timescales related to the parameter $\alpha$ and $\alpha^2$ are given respectively by:
\begin{equation*}
\tau = \frac{1+\alpha}{1-\alpha}, \qquad \tau' = \frac{1+\alpha^2}{1-\alpha^2} = \frac{\tau}{2} + \frac{1}{2\tau}.
\end{equation*}
We now rewrite the above result with the new filter $\mathcal{L}_\tau$ and the parameter $\tau$ instead of $\mathcal{F}_\alpha$ and parameter $\alpha$: $\mathcal{F}_\alpha = (\tau + 1)\mathcal{L}_{\tau}/2$,  $\mathcal{F}_{\alpha^2} = (\tau + 1)^2\mathcal{L}_{\tau'}/4\tau$. Insert these expressions into the result of the above theorem, we obtain the equation for discrete EMA filter:
\begin{equation}\label{filter}
\left(1-\frac{1}{\tau}\right)\mathcal L_{\tau'} \Big[X_t\mathcal L_\tau [X_{t-1}]\Big] = \mathcal L_\tau^2 [X_t]  - \frac{1}{\tau}\mathcal L_{\tau'}\big[X_t ^2\big].
\end{equation}

%

\subsection{Generalized trend formula for trend-following}
\subsubsection{Trend formula in continuous-time framework}
In order to derive the generalized formula, we employ the continuous-time approach framework. Within this framework, we consider the stochastic process of the asset price $S_t$. The trend of the asset $P_t$ can be obtained by using the the linear filter in continuous-time described by the following stochastic differential equation (SDE):
\begin{align} \label{eq:Kalman_filter}
d P_t = -\frac{2}{\tau}P_t d t + \frac{2}{\tau}dS_t.
\end{align}
The solution of this SDE given by the following expression:
\begin{equation}
P_t = \frac{2}{\tau}\int_{-\infty}^t e^{2(t-s)/\tau} dS_s.
\end{equation}
In this form, the above filter can be also interpreted as the exponential weight moving average filter $\mathcal{L}_\tau$ in the discrete-time framework. Hence, we employ the notation $\mathcal{L}_\tau$ to rewrite the above filter equation $P_t=\mathcal{L}_\tau[S_t]$. Building the trend-following strategy using this trend estimation deformed by a non-linear function  $\phi(x)$, we obtain the of the change of profit and loss ($G_t$):   
\begin{equation}\label{eq:pnl_eq}
d G_t = \phi(P_t) \times d S_t.
\end{equation}
Here $\phi(x)$ can be a function like $\phi(x) = \mathrm{sign}(x)$ or $\phi(x)= \mathrm{Cap}_{\Lambda_1,\Lambda_2}(x)$ in order to limit the extreme exposure to the asset. \medskip

Using the Kalman filter (Eq. \ref{eq:Kalman_filter}) to eliminate the dependence of the change of profit and loss (Eq. \ref{eq:pnl_eq}) on the asset price $S_t$, we obtain:
\begin{align*} 
d G_t & = \phi(P_t)P_t d t + \frac{\tau}{2}\phi(P_t) d P_t.	
\end{align*}

Let $F(x)$ be such that $F'(x)=\phi(x)$, then using Ito's lemma we have:
\begin{equation*}
d F(P_t) = \phi(P_t)dP_t + \frac{2\phi'(P_t)}{\tau^2} d S_t^2.
\end{equation*}
Inserting this expression in the equation of P\&L, we obtain:
\begin{equation*}
d G_t =  \frac{\tau}{2} d F(P_t) + \phi(P_t)P_t d t -  \frac{\phi'(P_t)}{\tau}dS_t^2.
\end{equation*}
We rearrange different terms of the P\&L equation and introduce new timescale $T$:
\begin{equation*}
d \left(\frac{\tau}{T}F(P_t)\right) =  - \frac{2}{T} \left(\frac{\tau}{T}F(P_t)\right)dt + \frac{2}{T}\left(d G_t
+  \frac{\phi'(P_t)}{\tau}dS_t^2 
- \phi(P_t)P_td t + \frac{\tau}{T}F(P_t)d t\right).
\end{equation*}
Writing the above equation in form of $\mathcal{L}_T$ filter, we obtain:
\begin{equation*}
\frac{\tau}{T}F(P_t) = \mathcal{L}_T \left[d G_t
+  \frac{\phi'(P_t)}{\tau}dS_t^2 
- \phi(P_t)P_td t + \frac{\tau}{T}F(P_t)d t\right].  
\end{equation*}
Finally, we derive the generalized equation for trend P\&L:
\begin{equation}\label{eq:generalized_trend}
\mathcal{L}_T [d G_t] = \frac{\tau}{T}F(P_t) - \mathcal{L}_T\left[  \frac{\phi'(P_t)}{\tau}dS_t^2\right]
+ \mathcal{L}_T \left[\phi(P_t)P_t - \frac{\tau}{T}F(P_t)\right] d t.
\end{equation}

%

\subsubsection{Linear trend estimation}
In the case where $\phi(x)=x$, we have $F(x) =x^2/2$, then we find the result showed for discrete approach: 
\begin{equation}
\mathcal{L}_T [d G_t] = \frac{\tau}{2T}P_t^2 - \frac{1}{\tau}\mathcal{L}_T[dS_t^2]
+ \Big(1-\frac{\tau}{2T}\Big)\mathcal{L}_T [P_t^2]d t.
\end{equation}
With the choice of timescale $T=\tau/2$ we eliminate the last term (correction term) then obtain:
\begin{equation}
\mathcal{L}_T [d G_t] = P_t^2 - \frac{1}{2T}\mathcal{L}_T[dS_t^2].
\end{equation}

\subsubsection{Non-linear trend estimation}\label{apdx:non_linear_trend}
Let us consider now the case $\phi(x)=\mathrm{sign}(x)$ hence $F(x)=|x|$ and its derivative is $\phi'(x)=2\delta(x)$. We obtain:
\begin{equation*}
\mathcal{L}_T [d G_t] = \frac{\tau}{T}|P_t| - \frac{2}{\tau}\mathcal{L}_T \Big[\delta(P_t)dS_t^2\Big] 
+ \Big(1-\frac{\tau}{T}\Big)\mathcal{L}_T [|P_t|]d t. 
\end{equation*}
With the choice of timescale $T=\tau$, we eliminate the last term (correction term) then obtain the following equation:
\begin{equation*}
\mathcal{L}_\tau [d G_t] = |P_t| - \frac{2}{\tau}\mathcal{L}_\tau \Big[\delta(\hat{\mu}_t)dS_t^2\Big].
\end{equation*}
For return $dS_t$ is risk managed at stable volatility $\sigma$ and follows Gaussian process $dS_t \sim \mathcal{N}(\bar{\mu},\sigma)$, we have the following approximation:
\begin{equation*}
\Big\langle \mathcal{L}_\tau \Big[\delta(P_t)dS_t^2\Big] \bgiven P \Big\rangle \approx \sigma^2 \langle \delta(P)\rangle.
\end{equation*}
As the trend estimate $P_t$ follows the following distribution $\mathcal{N}(\bar{\mu},\sigma/\sqrt{\tau})$, we have:
\begin{equation*}
\langle \delta(P)\rangle = \int_{-\infty}^{\infty} \delta(x) \frac{1}{\sqrt{2\pi}\sigma_\mu}e^{-x^2/2\sigma_\mu} dx = \sqrt{\frac{\tau}{2\pi}}\frac{1}{\sigma}.
\end{equation*} 
Inserting this approximation in the P\&L equation, we obtain the result for the case of sign of trend:
\begin{equation}
\langle\mathcal{L}_\tau [d G_t] \given P \rangle = |P| - \sqrt{\frac{2}{\pi\tau}}\sigma.
\end{equation}

\end{document}